\documentclass{kapproc}
\usepackage{t1enc}
\usepackage{procps} 
\usepackage[dvips]{graphicx}
\upperandlowercase
\kluwerbib

\def\spose#1{\hbox to 0pt{#1\hss}}
\def\lta{\mathrel{\spose{\lower 3pt\hbox{$\mathchar"218$}}
     \raise 2.0pt\hbox{$\mathchar"13C$}}}
\def\gta{\mathrel{\spose{\lower 3pt\hbox{$\mathchar"218$}}
     \raise 2.0pt\hbox{$\mathchar"13E$}}}
\def\etal{{\it et al.\ }}

\begin{document}

\articletitle{Star Formation Efficiencies and \\
Star Cluster Formation}


\author{Uta Fritze -- v. Alvensleben}
\affil{Universit\"atssternwarte G\"ottingen\\
Geismarlandstr. 11, 37083 G\"ottingen, Germany}
\email{ufritze@uni-sw.gwdg.de}


\begin{abstract}
Starbursts produce large numbers of {\bf Y}oung {\bf S}tar 
{\bf C}lusters ({\bf YSC}s). Multi-color photometry in combination with a dedicated SED analysis 
tool allows to derive ages, metallicities, ${\rm E_{B-V}}$, and masses including $1 \sigma$ uncertainties for individual clusters and, hence, mass 
functions for YSC systems. 
The mass function, known to be Gaussian for old 
{\bf G}lobular {\bf C}luster ({\bf GC}) systems, is still controversial for 
YSC systems. GC formation is expected 
in massive gas-rich spiral -- spiral mergers because of their high global star formation 
efficiencies and observed in $\geq 1$ Gyr old merger remnants. Yet it has not been possible to identify young GCs among 
YSC populations. We suggest a compactness parameters involving masses and 
half-light radii of YSCs to 
investigate if young GCs are formed in starbursts and if the ratio of young GCs to more loosely bound star clusters 
depends on galaxy type, mass, burst strength, etc.
\end{abstract}

\section{Star Formation Efficiencies \& Star Cluster Formation}
Both burst strengths b defined by the relative increase of the stellar mass
in the course of starbursts, ${\rm b:=\frac{\Delta S_{burst}}{S_{total}}}$, and
Star Formation Efficiencies (SFEs) defined as the
total stellar mass formed out of an available mass of gas, ${\rm SFE :=}$ 
${\rm \frac{\Delta S_{burst}}{G}}$, are difficult to determine. Reasonable estimates are only possible in young post-starbursts. As
long as a burst is active only lower limits can be given. Once a burst is over 
or if a burst lasts longer than the most massive stars' lifetimes, the amount
of stars already died needs to be accounted for. For 
the stellar and gaseous masses before the burst can only be estimated on the basis of Hubble types, HI observations, etc. 
The strongest bursts are reported in mergers of
massive gas-rich galaxies with total burst durations of the order of a few
100 Myr. 
Bursts in massive interacting galaxies are much stronger and last much longer
than those in isolated dwarf galaxies. Blue Compact Dwarf
galaxies (BCDs), e.g., feature bursts with durations of the order of a 
few Myr, ${\rm b \ll 0.1}$, ${\rm SFE \leq 0.01}$, and a trend of decreasing 
burst strengths for increasing total galaxy masses (including HI) 
(Kr\"uger \etal 1995). Massive interacting galaxies feature bursts 
stronger and more efficient by one to two
orders of magnitude, similar to the
progenitors of E+A galaxies in clusters, ULIRGS, SCUBA galaxies, and
optically identified starburst galaxies in the early universe. 
The post-burst spiral -- spiral merger remnant NGC 7252 with 
two long, gas-rich tidal tails pointing at an age of $\lta 1$ Gyr after the
onset of the strong interaction and its blue and radially constant colors and very strong Balmer absorption line spectrum must have experienced a very strong and global starburst
increasing its stellar mass by as much as $\sim 40$ \%  
between 600 and 1000 Myr ago. Conservative estimates still lead to a very high
SFE $\geq 30$\% (Fritze -- v. Alvensleben  \& Gerhard 1994). A large number of Star Clusters (SCs) 
formed throughout the main body,
many of them apparently so strongly bound that they managed to survive for 
500 $-$ 900 Myr the violent relaxation phase that restructured the remnant into a de Vaucouleurs profile (cf. Fritze -- v. Alvensleben \& Burkert 1995 and 
Schweizer 2002 for a recent review).
Most of these star clusters are young Globular Clusters (GCs) 
based on their ages, luminosities, and radii. How many clusters were
already destroyed since the onset of the burst? 
An analogous system at a younger age is NGC 4038/39 where the two spiral
disks just started overlapping. Its burst around the two nuclei,
along the tidal structures, and -- strongest -- in the optically obscured
disk -- disk overlap region is in its initial stage. Thousands of bright 
Young Star Clusters (YSCs) are
seen with luminosities ranging from those of individual red supergiant stars to
${\rm M_V \geq -15}$. How many of these will survive for $\gg 1$ Gyr
and become Globular Clusters (GCs)? 

Hydrodynamic modelling shows that the formation of long-lived strongly bound GCs
requires SFEs $\gg 10$ \%, originally thought
to only occur in the early universe. In normal SF in spirals,
irregulars, and starbursting isolated dwarfs like BCDs, GC formation should
not be possible. In the high pressure ISM with its strong shocks in spiral
-- spiral mergers, however, GC formation is observed in reality  
and in high-resolution hydrodynamical models (Yuexing \etal 2004). Young
and intermediate age GCs hence are tracers of high SFE periods in their 
parent galaxies. 
A number of very fundamental questions are still open at present: 
Does the amount of SF that goes into massive, compact, long-lived SCs scale with
burst strength and/or (local/global) SFE? Or is there a threshold in SFE, below
which only field stars and weakly bound, less massive SCs or OB-associations can
be formed that dissolve on timescales of $10^8$ yr, and above which GCs can be
formed or even become the dominant component? Does the same star and SC formation
mechanism work in vastly different environments and scale over a huge dynamical
range or are there two different modes of SF like ``normal'' and ``violent''? 
SCs are seen to form in many environments from normal Irrs and
spirals through dwarf starbursts, spiral mergers, and ULIRGs, constrained to nuclear regions (e.g. in ULIRGs), over their main body (e.g. NGC 4038/39), and all along tremendous tidal tails (e.g. 
Tadpole cf. de Grijs \etal 2003). The spatial extent of a starburst probably depends on the orbit and relative
orientations of the interacting galaxies, on whether or not they had massive 
bulges and/or DM halos. Are all
these YSC systems similar or systematically different in terms of
masses, mass functions, sizes, compactness or degree of binding and, hence,
survival times.

SCs are {\bf S}imple {\bf S}tellar {\bf P}opulations ({\bf SSP}s) with all stars
having the same age and chemical composition. Evolutionary synthesis models 
like {\bf GALEV} describe the evolution of
SCs over a Hubble time, from the youngest stages of 4 Myr all through the oldest
GC ages $\geq 14$ Gyr for 5 different
metallicities ${\rm -1.7 \leq [Fe/H] \leq}$ $+0.4$. The TP-AGB phase is very important for age-dating of SCs between 100 Myr and a few Gyr on the basis of their ${\rm
V-I}$ colors
(cf. Schulz \etal 2002). Gaseous emission in terms of 
lines and continuum for the respective metallicities makes important
contributions to broad band fluxes and colors at young ages (Anders \& Fritze -- v. Alvensleben 2003). Lick absorption indices significantly help disentangle ages and
metallicities of older SCs (Lilly \& Fritze -- v. Alvensleben {\sl in prep.}). {\bf GALEV} models yield the
detailed spectral evolution of SCs from 90 \AA \ through 160 $\mu$m, luminosities,
M/L-ratios, and colors in many filter systems (Johnson, HST, Washington,
Stroemgren, \dots) and can be retrieved from
{\sl http://www.uni-sw.gwdg.de/$^{\sim}$galev/}\ . 

\section{Analysing Star Cluster Systems}
The time evolution of luminosities, colors, and M/L-ratios significantly depends
on metallicity in a way that is different in different wavelengths regions. For young
SC systems, like in NGC 4038/39, extinction is an important issue. Older
starbursts, like in NGC 7252, are significantly less extincted. 
An ESO -- ASTROVIRTEL project provides us with HST and VLT multi-$\lambda$ photometry for SC
systems from young to old that have 4 and more passbands observed. A dedicated Spectral Energy 
Distribution (SED) analysis tool called AnalySED compares observed SC SEDs with an extensive grid
of $117 \, 000$ SSP model SEDs for 5 different metallicities, 
1170 ages from 4 Myr through 14
Gyr, and 20 extinction values ${\rm 0 \leq E_{B-V} \leq 1}$. 
We use Calzetti \etal 's (2000) starburst extinction law since internal 
extinction is only an issue in ongoing starbursts.
A probability ${\rm p(n) \sim exp(-\chi^2)}$ is assigned to each model SED 
by a
maximum likelihood estimator ${\rm \chi^2=\sum_{\lambda}(m_{\lambda}^{obs}-
m_{\lambda}^{model})^2/\sigma_{obs}^2}$. The best fit model is the one with the
highest probability. Probabilities are normalised to ${\rm \sum_n p(n)=1}$.
Summing models with decreasing probabilities until ${\rm \sum_n p(n)=0.68}$
provides $\pm 1 \sigma$ uncertainties for ages, metallicities, exticntion values, and masses of individual SCs (Anders \etal 2004a). 
Testing AnalySED with artificial SCs, we found 
that there are good and bad passband combinations,
slightly depending on the ages and metallicities of the clusters, and we
identified a combination of 4 passbands U, B, V or I, and H or K with 
observational photometric accuracy $\leq 0.2$ mag as optimal for YSC 
systems. We agree with the independent investigation by Cardiel 
\etal (2003), that at typical photometric accuracies broad band photometry
with useful passband combinations is as powerful in disentangling ages and
metallicities as is spectroscopy with typical S/N. 
The AnalySED tool is currently extended to also include Lick indices for
analyses of intermediate-age and old GC systems (Lilly \etal, {\sl in prep.}, 
cf. Lilly's poster on the accompanying CD-ROM). 
In the dwarf starburst galaxy NGC 1569 we identify 169 YSCs on the ASTROVIRTEL
images, the bulk of them with ages $\leq 25$ Myr, low extinction and metallicities. Their
masses are typically in the range from $10^3$ to ${\rm 10^4 M_{\odot}}$, 
only the two Super SCs have 
masses in the mass range of GCs (cf. Fig.1 and Anders \etal 2004b).

\begin{figure}[ht]
\begin{center}
\includegraphics[width=6.cm]{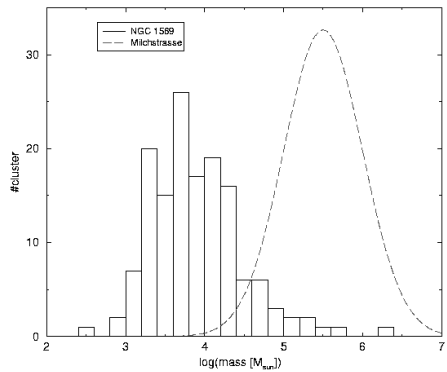}
\end{center}
\caption{Distribution of YSC masses in NGC 1569 (histogram) as compared to the
Milky Way GC mass function, normalised to the same number of clusters (Gaussian
curve).} 
\end{figure}

We conclude that the starburst in the dwarf galaxy NGC 1569 did not form
many GCs, whereas the starburst in the spiral -- spiral merger NGC 7252 did. 
Does SC formation produce a continuum in masses and binding energies or are
there different modes of SC formation that respectively produce open and
globular clusters? With increasing burst strength an increasing number
of SCs is formed. Does the statistical effect of having a higher chance to have
a more massive cluster within a larger sample explain the difference
between cluster masses in NGC 1569 and NGC 7252? Probably not. Mass Functions (MFs) are power laws for open
cluster systems and Gaussians for  old 
GC systems. Initial MFs derived from models for survival 
and destruction of GCs in galactic potentials are controversial. 
Vesperini (2001) 
favors an initially Gaussian shape for GC MFs that is essentially conserved
by the competing destructions of low-mass GCs through tidal disruption and of high-mass GC by
dynamical friction. Gnedin \& Ostriker (1997) favor an initially power-law GC MF
that is secularly transformed into a Gaussian by higher destruction rates for
lower mass GCs. The MF of the young SC system in NGC 4038/39, that very probably
comprises a mixture of OB-associations, open, and GCs, as derived 
from HST photometry, is also controversial. Whereas Zhang \& Fall (1999)
derive a power-law MF using reddening-free Q$_1$Q$_2$ indices, we find a
Gaussian MF (Fritze -- v. Alvensleben 1998, 1999). Both approaches have
their drawbacks. Zhang \& Fall excluded a
significant number of clusters from the ambiguous age range in the Q$_1~-~$Q$_2$
- plot, we assumed a uniform average reddening in the WFPC1 UVI data. If we exclude the same SCs as
Zhang \& Fall, we also find a power-law. The dust distribution in NGC
4038/39 is clearly patchy and a reanalysis of ASTROVIRTEL UBVIK data
from HST WFPC2 and VLT ISAAC underway (Anders \etal {\sl in prep.}). 
The shape of the MF need not correspond to the shape
of the Luminosity Function (LF) for a young SC system, 
as age spread effects can distort the shape of the LF with respect to that of the
underlying MF -- to the point of transforming a Gaussian MF into a
power-law LF up to the observational completeness limit. 
The key to survival or destruction is the strength of a SC's 
internal gravitational binding, as measured for Galactic GCs by their concentration parameters c. By definition, ${\rm c:= log
\frac{r_t}{r_c}}$ involves the tidal and core radii. Very young clusters need
not be tidally truncated yet and tidal radii could not be measured for the bulk of the 
YSCs on top of the bright galaxy background in NGC 4038/39 anyway. 
We therefore define the compactness of a young SC by the ratio between its mass
and half-light radius (cf. Anders \etal {\sl in prep.}), a robust quantity that can reliably be measured and is
predicted by dynamical SC evolution models not to significantly change over 
a Hubble time. To this aim, we first have to
improve upon the determination of SC radii by using appropriate aperture
corrections. Improved SC radii, in turn, lead to improved SC photometry, and,
hence, to improved photometric masses 
(cf. Poster by P. Anders on the accompanying CD-ROM). 

\section{Conclusions and Perspective}
From the ages, masses, and radii of their SCs we know that major gas-rich mergers can form
significant secondary populations of GCs in their strong and global starbursts.
SFEs in mergers are higher by 1$-$2 orders of magnitude
than in normal SFing galaxies and (non-interacting) dwarf galaxy starbursts. Comparing good precision
photometry in at least 4 reasonably chosen passbands (e.g. UBVK) to GALEV
evolutionary synthesis models for SCs with given age, metallicity, extinction,
and mass by means of a dedicated SED analysis tool (AnalySED) allows to
reliably determine individual SC ages, metallicities, extinction values, and
masses, including their respective 1$\sigma$ uncertainties. The first dwarf
galaxy starburst analysed in detail this way shows only very few clusters
with masses in the range of GCs among its $\sim 170$ YSCs. Clearly, both more 
major merger and dwarf galaxy starbursts need to be analysed in detail.
Pixel-by-pixel analyses (de Grijs \etal 2003) or integrated field spectroscopy can provide burst strengths and SFEs. From a comparison with 
HST multi-$\lambda$
imaging of their YSC systems the relative ratios of SF going into
field stars, short-lived open clusters, and long-lived GCs, respectively, can be determined. A key
question is whether these quantities as
well as the intrinsic properties of the YSCs, like masses and half-mass radii,
depend on environment or not, in a smooth way or with some threshold. A
comparison of starbursts in dwarf and massive, interacting and non-interacting
starbursts should tell if SF and SC formation are universal processes or
depend on environment. GC age and metallicity distributions will allow to trace back a galaxy's violent (star) formation
history and constrain galaxy formation scenarios (Fritze --
v. Alvensleben 2004). This requires B through
NIR photometry and medium resolution spectra to measure Lick indices.
With only one observed color we cannot disentangle the age -- metallicity degeneracy of
intermediate-age and old stellar populations 
and see if more than one GC population is hidden in
the red peak of many elliptical galaxies' bimodal color distributions. 

\begin{acknowledgments} I gratefully acknowledge travel support, in part from the DFG (FR 916/10-2) and in part from the organisers.\end{acknowledgments}

\end{document}